\documentclass[useAMS,usenatbib]{mn2e}

\usepackage{epsfig}
\usepackage{amsmath}
\usepackage{amssymb}
\usepackage{natbib}
\usepackage{threeparttable} 
\usepackage{booktabs}
\usepackage{float}
\usepackage{times}

\usepackage{epstopdf}

\newcommand{\chan}{\textit{Chandra}}
\newcommand{\swift}{\textit{Swift}}
\newcommand{\rxte}{\textit{RXTE}}
\newcommand{\xmm}{\textit{XMM-Newton}}

\newcommand{\maxi}{\textit{MAXI}}

\newcommand{\gaia}{\textit{Gaia}}

\newcommand{\Msun}{\mathrm{M}_{\odot}}
\newcommand{\lum}{\mathrm{erg~s}^{-1}}
\newcommand{\flux}{\mathrm{erg~cm}^{-2}~\mathrm{s}^{-1}}

\newcommand{\cnts}{\mathrm{c~s}^{-1}}
\newcommand{\mdot}{\mathrm{M_{\odot}~yr}^{-1}}

\newcommand{\dens}{\mathrm{g~cm}^{-3}}
\newcommand{\nh}{\mathrm{cm}^{-2}}
\newcommand{\dist}{(D/5.0~\mathrm{kpc})^2}

\newcommand{\source}{Aql X-1}

\newcommand{\ks}{KS~1731--260}
\newcommand{\mxb}{MXB~1659--29}

\newcommand{\maxisource}{MAXI~J0556--332}

\newcommand{\rxs}{1RXS J180408.9--342058}

\def \mnras {MNRAS}
\def \apj {ApJ}
\def \apjs {ApJS}
\def \apjl {ApJL}
\def \aap {A\&A}
\def \nat {Nature}

\def \atel {ATel}

\def \pasj {PASJ}

\def \prc {PhRvC}

\def \pre {PhRvE}

\def \iaucirc {IAU Circ.}

\def \aapr {A\&ARv}

\title[Crust cooling in \source]{Crust cooling of the neutron star in Aql X-1: Different depth and magnitude of shallow heating during similar accretion outbursts}
\author[N. Degenaar et al.]
{N. Degenaar$^1$\thanks{e-mail: degenaar@uva.nl}, L. S. Ootes$^{1}$, D.~Page$^{2}$, R. Wijnands$^{1}$, A.S. Parikh$^{1}$, J. Homan$^{3,4}$,  
\newauthor E. M. Cackett$^{5}$, J. M.~Miller$^{6}$, D.~Altamirano$^{7}$, and M.~Linares$^{8}$\\
$^1$Anton Pannekoek Institute for Astronomy, University of Amsterdam, Postbus 94249, 1090 GE Amsterdam, The Netherlands\\
$^2$Instituto de Astronom\'{i}a, Universidad Nacional Aut\'{o}noma de M\'{e}xico, Mexico D.F. 04510, Mexico\\
$^3$Eureka Scientific, Inc., 2452 Delmer Street, Oakland, CA 94602, USA\\
$^4$SRON, Netherlands Institute for Space Research, Sorbonnelaan 2, 3584 CA Utrecht, The Netherlands\\
$^5$Department of Physics and Astronomy, Wayne State University, 666 W. Hancock St, Detroit, MI 48201, USA\\
$^6$Department of Astronomy, University of Michigan, 500 Church Street, Ann Arbor, MI 48109, USA\\
$^7$Department of Physics and Astronomy, Southampton University, Southampton SO17 1BJ, UK\\
$^8$Departament de F\'{i}sica, EEBE, Universitat Polit\`{e}cnica de Catalunya, c/ Eduard Maristany 10, E-08019 Barcelona, Spain
}

\begin{document}

\date{Draft version}

\pagerange{\pageref{firstpage}--\pageref{lastpage}} \pubyear{0000}

\maketitle

\label{firstpage}

\begin{abstract}
The structure and composition of the crust of neutron stars plays an important role in their thermal and magnetic evolution, hence in setting their observational properties. One way to study the properties of the crust of a neutron star, is to measure how it cools after it has been heated during an accretion outburst in a low-mass X-ray binary (LMXB). Such studies have shown that there is a tantalizing source of heat, of currently unknown origin, that is located in the outer layers of the crust and has a strength that varies between different sources and different outbursts. With the aim of understanding the mechanism behind this ``shallow heating'', we present \chan\ and \swift\ observations of the neutron star LMXB Aql X-1, obtained after its bright 2016 outburst. We find that the neutron star temperature was initially much lower, and started to decrease at much later time, than observed after the 2013 outburst of the source, despite the fact that the properties of the two outbursts were very similar. Comparing our data to thermal evolution simulations, we infer that the depth and magnitude of shallow heating must have been much larger during the 2016 outburst than during the 2013 one. This implies that basic neutron star parameters that remain unchanged between outbursts, do not play a strong role in shallow heating. Furthermore, it suggests that outbursts with a similar accretion morphology can give rise to very different shallow heating. We also discuss alternative explanations for the observed difference in quiescent evolution after the 2016 outburst.
\end{abstract}

\begin{keywords}
accretion, accretion disks -- dense matter -- stars: neutron -- X-rays: binaries -- X-rays: individual (\source)
\end{keywords}


\section{Introduction}\label{sec:intro}
The liquid, dense core of neutron stars is covered by a $\sim$1-km thick, solid crust. The structure and composition of the crust of neutron stars play a key role in the evolution of their magnetic field strength and interior temperature. As such, the crust properties are important for a variety of observational phenomena such as, for instance, pulsar glitches, thermonuclear X-ray bursts, and magnetar outbursts, as well as gravitational wave signals from neutron stars \citep[e.g.][]{brown1998,ushomirsky2000,cumming2001_bursts,horowitz2009_gravwaves,pons2009,page2013}. This provides a strong incentive to understand the detailed properties of the crust of neutron stars.

Neutron stars that are part of a binary system in which they can accrete gas from a companion star with a mass $\lesssim 1~\Msun$, are called low-mass X-ray binaries (LMXBs).  Many neutron star LMXBs are transient and spend most of their time in quiescence. The gas that is transferred from the companion is then only accreted at high rates during weeks--years long outbursts. 
During such outbursts, LMXBs can exhibit different ``spectral states'' that are characterized by specific X-ray spectral and fast-variability properties \citep[][]{hasinger1989}, and likely reflect different accretion geometries \citep[e.g.][for a review]{done2007}.

The gas that accretes on to the surface of neutron stars undergoes thermonuclear burning, which transfers light elements (i.e. the accreted H and/or He) into heavier ones \citep[e.g.][]{schatz1999}. This thermonuclear burning is often unstable and causes run-away energy production \citep[e.g.][]{wallace1981} that results in a  thermonuclear X-ray burst \citep[X-ray burst hereafter; e.g.][for reviews]{lewin95,galloway2017}. 

Apart from causing thermonuclear burning on the surface, accretion fires up a series of nuclear reactions in the crust of neutron stars. This includes electron captures by nuclei in the outer layers of the crust, and density-driven fusion of nuclei at several hundreds meter depth. Based on theoretical calculations \citep[e.g.][]{haensel1990a,yakovlev2006,fantina2018}, and laboratory data from nuclear experiments \citep[e.g.][]{gupta07,estrade2011}, the energy produced in these nuclear reactions is thought to be $\sim$2~MeV per accreted nucleon \citep[e.g.][]{haensel2008}. This energy release can significantly heat the crust and bring it out of thermal equilibrium with the core. Since most energy is generated in the nuclear fusion reactions, this process is referred to as ``deep crustal heating" \citep[][]{brown1998}. The crust cools during quiescent phases, when the heat energy gained during outburst is thermally conducted throughout the neutron star \citep[e.g.][]{colpi2001,rutledge2002,brown08,page2012,wijnands2012}.

Over the past decade, evidence has been accumulating that during accretion phases in LMXBs, the crust of neutron stars are more strongly heated than predicted by nuclear heating models. This inference comes from observations of different types of phenomena. For instance, observations of very long X-ray bursts and mHz quasi-periodic oscillations (QPOs) point to higher crust temperatures than can be accounted for with standard nuclear heating \citep[e.g.,][]{cumming06,keek2008_1608,altamirano2012,zand2012,linares2012_ter5_2}. Furthermore, detailed  monitoring of the quiescent temperature evolution of neutron stars following accretion outbursts, has highlighted that there is something missing in our understanding of how neutron star crusts are heated \citep[e.g.][]{brown08,degenaar2011_terzan5_3}.

Ten neutron stars have been monitored after accretion outbursts with the aim to study how their accretion-heated crusts cool in quiescence \citep[see][for a review]{wijnands2017}. The high temperatures observed within the first few hundred days of quiescence 
require an additional source of energy to heat the crust \citep[e.g.][]{brown08,degenaar2011_terzan5_3}. 
The strength of this ``shallow heating'', if proportional to the mass-accretion rate, is inferred to be on the order of $\sim$1--2~MeV~nucleon$^{-1}$ \citep[e.g.][]{degenaar2011_terzan5_3,degenaar2014_exo3,degenaar2015_ter5x3,page2013,parikh2017,parikh2018,ootes2018}, although one extreme case requires $\sim$15--17~MeV~nucleon$^{-1}$  \citep[e.g.][]{homan2014,deibel2015,parikh2017_maxi}. It is currently unclear what is causing this shallow heating. 

It remains to be established to what extent shallow heating depends on neutron-star specific properties (e.g., spin, magnetic field strength, mass, radius, superfluid properties, age), and on the detailed properties of the accretion outburst (e.g., brightness, duration, accretion geometry). Studying multiple cooling curves of a single source may be a promising way to understand the origin of shallow heating; because the fundamental properties of a neutron star remain virtually unchanged, the effect of the outburst parameters on the crust heating and cooling can potentially be isolated \citep[][]{waterhouse2016,parikh2017_maxi,parikh2018}. 

\subsection{The frequently active neutron star LMXB Aql X-1} 
Aql X-1 is a transient neutron star LMXB that was discovered in the late sixties \citep[][]{friedman1967} and has been seen active many times since \citep[e.g.][]{kaluzienski1977,kitamoto1993,campana2013,gungor2014_aqlx1,ootes2018}. Its accretion outbursts last $\sim$1--6 months, vary in brightness from $L_{\mathrm{X}}$$\simeq$$10^{35}$ to $10^{38}~\dist~\lum$, and recur on a timescale of $\sim$1 yr. The neutron star displays X-ray bursts \citep[e.g.][]{koyama1981}, and spins at a frequency of $\simeq$550~Hz \citep[][]{zhang1998_bo,casella2008}. The source is located at $\approx 5$~kpc \citep[e.g.][]{rutledge2001,galloway06}\footnote{We note that Aql X-1 appears in \gaia\ DR2, but due to its faintness the parallax error is large and therefore the inferred distance estimate becomes strongly dependent on the assumed prior.}, and the binary companion is a K-type star that orbits the neutron star in $\approx$19 hr \citep{callanan1999,matasanchez2017_aqlx1}. 

With its frequent outbursts of varying length and brightness, Aql X-1 could be a promising source to gain more insight into shallow crustal heating \citep[][]{waterhouse2016}.  However, the origin of its quiescent X-ray emission has been debated. Firstly, apart from soft thermal emission from the neutron star surface, its quiescent X-ray spectrum often contains a hard emission tail. This component can be modeled as a $\Gamma \sim 1-2$ power law, and can contribute up to $\sim$80\% of the total unabsorbed 0.5--10 keV flux \citep[e.g.][]{verbunt1994,rutledge2002_aqlX1,campana2003_aqlx1,cackett2011_aqlx1,campana2014}. 
Such a power-law component is often seen in the spectra of quiescent LMXBs and could point to ongoing accretion \citep[see e.g.][for recent discussions]{chakrabarty2014_cenx4,dangelo2014,wijnands2015}. Secondly, days-long flares have been seen from \source\ during which the quiescent X-ray emission increased by a factor of $\gtrsim$10 \citep[][]{cotizelati2014}. It is commonly assumed that such flares are caused by a (short-lived) spurt of accretion \citep[e.g.][]{degenaar09_gc,fridriksson2011,wijnands2013}. 

Despite indications that there is at least some level of quiescent accretion activity in Aql X-1, \citet{waterhouse2016} argued that the crust of the neutron star may be so hot that its cooling drives the long-term quiescent flux evolution  \citep[because the injection of heat from continued low-level accretion is then very small compared to the heat content of the crust; see also e.g.][]{fridriksson2011,turlione2013,homan2014,bahramian2015,deibel2015,parikh2017}. \citet{waterhouse2016} studied X-ray data from the Neil Gehrels Swift observatory \citep[\swift;][]{gehrels2004} after three different outbursts of Aql X-1 (2011, 2013 and 2015). The quiescent spectra were dominated by a thermal emission component and the inferred temperatures were compared to a neutron star thermal evolution code \citep[\textsc{DStar};][]{brown08}. It was shown that the measured temperatures and decay trends could naturally be explained within the crust cooling paradigm \citep[][]{waterhouse2016}. 

The possibility of observing crust cooling in Aql X-1 was further investigated by \citet{ootes2018}, who tracked the evolution of the neutron star temperature over its 1996--2015 outburst history, using the thermal evolution code \textsc{NSCool} \citep[][]{page2016_nscool}, and allowing it to evolve both in quiescence and in outburst \citep[as implemented in \textsc{NSCool} by][]{ootes2016}. The data could be reproduced if the depth and magnitude of shallow heating were allowed to vary between different outbursts, although no clear correlation was found between the shallow heating parameters and the outburst properties. Moreover, it was found that the crust requires $\sim$1500~days to fully cool, which is longer than the recurrence time of the outbursts in Aql X-1 \citep[][]{ootes2018}. 

A new bright outburst from \source\ was detected in 2016 July \citep[e.g.][]{sanna2016_aqlx1_xrt}. Here we report on \chan\ and \swift\ observations of \source\ obtained after this bright outburst ceased. This study was aimed as a further test of the crust cooling hypothesis for the quiescent X-ray emission of Aql X-1, and to constrain the properties of shallow heating based on that assumption.


\section{Observations and data reduction}\label{sec:obs}

\subsection{Chandra observations}\label{subsec:chan}
As part of our \chan\ ToO program, Aql X-1 was observed at $\sim$50, 150, and 250~days after the end of its 2016 outburst. For all these observations the source was placed on the ACIS-S chip, using a 1/8 sub-array and the faint, timed data mode. 

The outburst ceased in 2016 September and our first \chan\ observation, lasting $\sim$14~ks, was obtained on November 22 starting at 15:57 \textsc{ut}. The second observation was performed on 2017 February 20, when it was observed for $\sim$18~ks starting at 12:43 \textsc{ut}. The third and final \chan\ observation was carried out on 2017 May 30, starting at 17:57 \textsc{ut}, and had an exposure time of $\sim$23~ks. Unfortunately, Aql X-1 had just gone into outburst at this time \citep[][]{vlasyuk2017_aqlx1}, after a relatively short quiescent phase of  $\sim$235~days. Since the X-rays in this third \chan\ observation will thus track the accretion emission and not that of the cooling neutron star, it cannot be used for our purposes and we will not discuss it further.

The \chan\ data were reduced within \textsc{ciao} (v. 4.9). A circular region with a radius of $1.5''$ was used to extract source events and a $15''$-radius circular region was used for the background. We used \textsc{dmextract} to extract count rates and light curves. During the first and second \chan\ observation, \source\ was detected at a net 0.3--10 keV count rate of $(9.45 \pm 0.26) \times 10^{-2}$ and $(5.92 \pm 0.21) \times 10^{-2}~\cnts$, respectively. No prominent variability is seen in the source light curves. We used \textsc{specextract} to extract spectra and to create response files. All spectral data were grouped to contain a minimum of 20 photons per bin using \textsc{grppha}.

\subsection{Swift observations}\label{subsec:swift}

\subsubsection{Outburst and quiescent XRT light curve}\label{subsec:swiftlc}
\begin{figure}
 \begin{center}
\includegraphics[width=8.5cm]{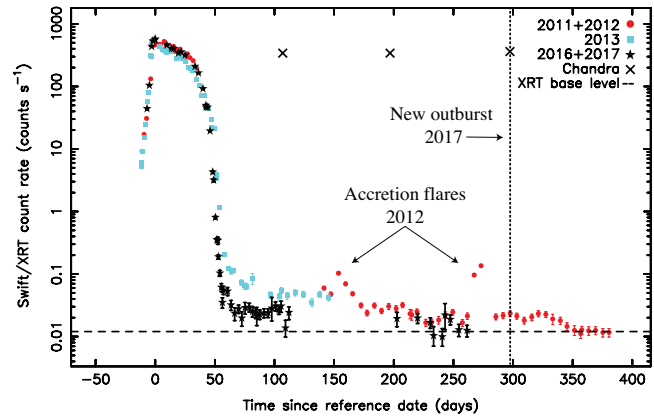}
    \end{center}
\caption[]{\swift/XRT (0.3--10 keV) light curves of the 2011 (red circles), 2013 (cyan squares) and 2016 (black stars) outbursts and subsequent quiescent phases of \source\ (binning is per observation). The light curves have been shifted such that the peak of the outburst corresponds to $t=0$. The times of our 2016/2017 \chan\ observations are indicated by the crosses in the top of the plot. The vertical dotted line indicates the time at which a new outburst started in 2017. For reference, the dashed horizontal line shows the XRT count rate previously identified as the quiescent based level \citep[][]{waterhouse2016}, although the true quiescent level of the source is likely lower \citep[][]{ootes2018}.
}
 \label{fig:lcxrt}
\end{figure}

Aql X-1 was monitored with \swift\ during its 2016 outburst and the subsequent quiescent phase, with the exclusion of a $\sim$3 month period during which the source was too close to the Sun. We used the \swift/XRT data performed in quiescence to obtain additional constraints on the temperature evolution of the neutron star. To determine which observations were suited for our quiescent analysis, we first produced an XRT count rate light curve. This was obtained using the online XRT repository\footnote{www.swift.ac.uk/user\_objects/} \citep[][]{evans2007,evans2009}, and is shown in Figure~\ref{fig:lcxrt}. 

\begin{figure}
 \begin{center}
\includegraphics[width=8.5cm]{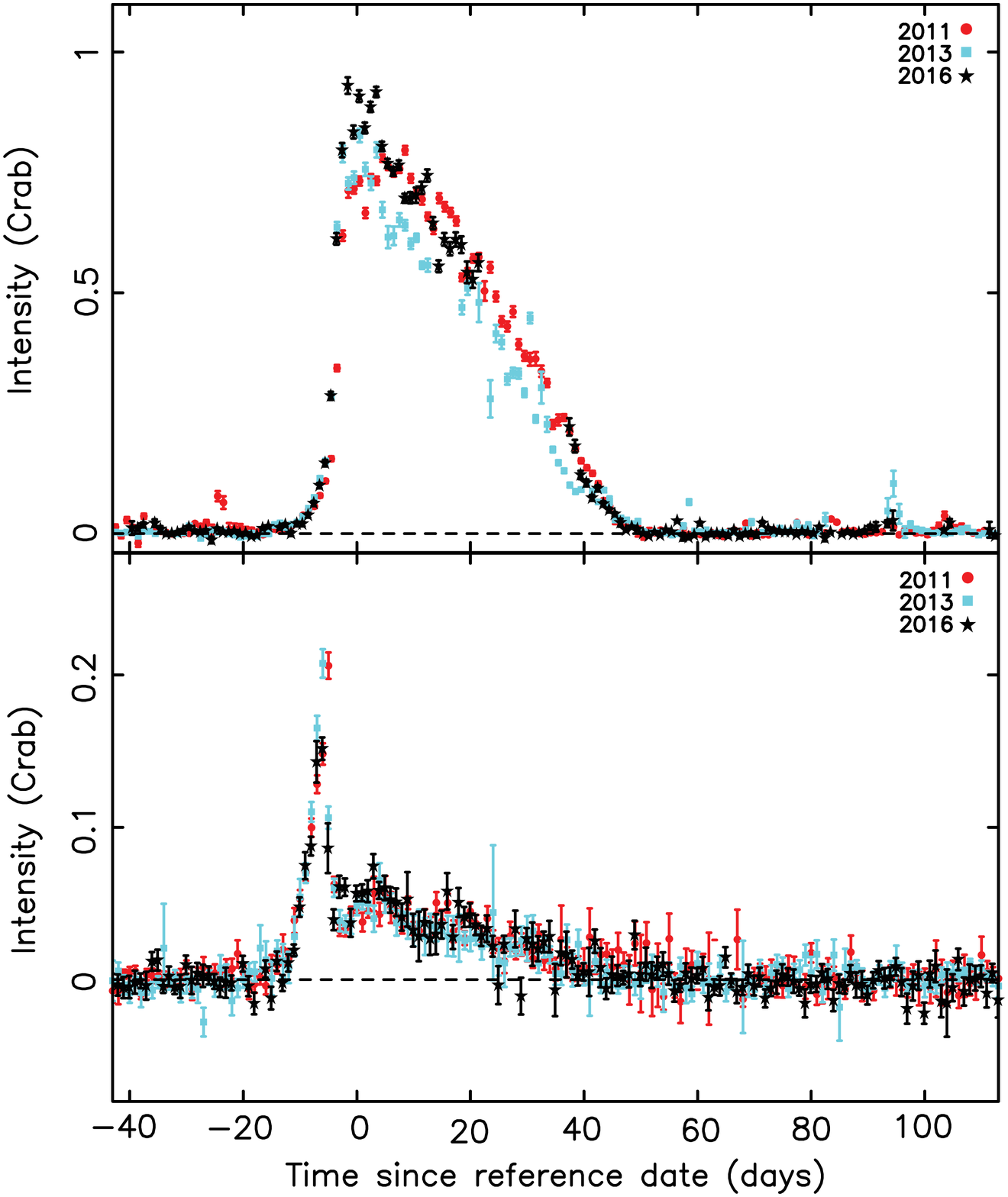} 
    \end{center}
\caption[]{\maxi\ (2--20 keV; top) and \swift/BAT (15--50 keV; bottom) light curves in Crab units, highlighting the 2011 (red circles), 2013 (cyan squares) and 2016 (black stars) outbursts of \source\ (binned per day). In both panels, the light curves have been shifted such that the 2--20 keV peak of the outburst corresponds to $t=0$. 
}
 \label{fig:lcmonit}
\end{figure}

One of our prime aims was to compare the quiescent evolution of Aql X-1 after its bright 2016 outburst with that observed after its bright 2011 and 2013 outbursts \citep[studied by][]{waterhouse2016,ootes2018}. Figure~\ref{fig:lcxrt} illustrates the striking resemblance between these three outbursts in terms of duration, peak flux, and overall shape. This is further illustrated by the \swift/BAT \citep[][]{krimm2013} and \maxi\ \citep[][]{maxi2009} monitoring light curves, which are shown in Figure~\ref{fig:lcmonit}. This suggest that the spectral state evolution was very similar during the three outbursts \citep[see also][]{diaztrigo2018}. 

The XRT light curve of the 2016 outburst shows a transition from a rapid decay that takes place over $\sim$2~weeks time (days $\sim$44--57 in Figure~\ref{fig:lcxrt}) to a much slower decay that continues until the new outburst commences. Such a prominent change in decay rate has been seen more often in neutron stars that were monitored for their crust cooling, and has been interpreted as the transition from the outburst decay to quiescence \citep[e.g.][]{fridriksson2010,homan2014,parikh2017_maxi,parikh2017}. As in \citet{waterhouse2016} and \citet{ootes2018}, we fitted the two different decay parts to exponential functions to estimate the onset of quiescence, $t_0$, as the intercept of the two. This yielded $t_0=$~MJD~57664 (2016 October 3), with exponential decay time scales of $\sim$1.7 and 145~days for the rapid and slow decay, respectively. This is similar to the results obtained for the 2013 outburst \citep[][]{waterhouse2016}.\footnote{No \swift/XRT observations were available during the decay of the 2011 outburst due to Sun constraints (see Figure~\ref{fig:lcxrt}).}

A first comparison of the quiescent evolution after the 2011, 2013 and 2016 outbursts is provided by the XRT light curves shown in Figure~\ref{fig:lcxrt}. It is of note that monitoring after the 2013 outburst stopped $\sim 100$~days into quiescence due to Sun-constraints, and for the same reason the decay and first $\sim 100$~days of quiescence were missed for the 2011 outburst \citep[][]{waterhouse2016}. Nevertheless, we can see from Figure~\ref{fig:lcxrt} that the XRT count rate light curve after the 2016 outburst deviates from that obtained after the 2013 outburst, i.e. up to $\sim 100$~days into quiescence. This provides a first hint that the early temperature evolution was not the same after (and perhaps during) the 2013 and 2016 outbursts. 

For the 2011 outburst we can only compare the late-time quiescent evolution, i.e. at $\gtrsim$100 days into quiescence. The first series of data points obtained after the 2011 outburst have no direct overlap with the 2016 data set, but the count rate is higher than the those measured shortly after the 2016 outburst. At $\sim$150~days in quiescence, when we can first directly compare the two data sets, the absolute count rates and evolution largely overlap (see Figure~\ref{fig:lcxrt}).

\subsubsection{Quiescent XRT spectra}\label{subsec:swiftspec}
To obtain temperature measurements of \source\ in quiescence in addition to those provided by our two \chan\ observations, we used 11 XRT observations performed between 2016 October 10 and November 7 (obsID 00034719016--26) and 10 observations performed between 2017 March 16 and May 10 (obsID 00033665090--99). The XRT light curve shown in Figure~\ref{fig:lcxrt} suggests that there were no accretion flares around the time of our \chan\ observations. 

Given the short exposures and low count rates (see Figure~\ref{fig:lcxrt}), the number of counts obtained for individual XRT observations was low ($\sim$10--30 counts). Therefore, we combined subsequent observations with similar count rates for our spectral extraction. To facilitate fits with a two-component model (see Section~\ref{subsec:qspec}), we aimed for a total of $\sim$100--200 counts per composite spectrum. This resulted in 3 different \swift/XRT epochs (see Table~\ref{tab:spec}).

All XRT observations that we used for our spectral analysis were obtained in PC mode. Reduction of these \swift\ data was performed within \textsc{heasoft} (v. 6.23). As an initial step, all observations were reprocessed with the \textsc{xrtpipeline}, using standard quality cuts. We next used \textsc{XSelect} to extract accumulated spectra. To obtain source spectra we used a circular extraction region with a radius of $30''$. Three circular regions with $30''$-radii, placed away from the source, were used to obtain background spectra. After summing the exposuremaps for the different observations in a particular epoch with \textsc{XImage}, an arf was created using \textsc{xrtmkarf}. The appropriate response matrix file (v. 14) was obtained from the calibration data base. All obtained spectra were grouped to contain a minimum of 15 photons per bin using \textsc{grppha}.

\subsubsection{Spectral fitting}\label{subsec:specfit}
All spectral fits were performed within \textsc{XSpec} \citep[v. 12.10;][]{xspec}. Based on previous quiescent studies of \source, we fitted the quiescent spectral data to a two-component model comprised of a neutron star atmosphere component \citep[\textsc{nsatmos};][]{heinke2007}, and a power-law component (\textsc{pegpwrlw}). We modeled the interstellar absorption with \textsc{tbabs}, using \textsc{vern} cross-sections and \textsc{wilm} abundances \citep[][]{verner1996,wilms2000}. 

As in \citet{waterhouse2016}, we fixed the \textsc{nsatmos}-model parameters $M=1.6~\Msun$, $R=11$~km, $D=5$~kpc, and $N=1$ (where the latter implies that the neutron star surface is uniformly radiating), and only left the neutron star temperature ($kT_{\mathrm{eff}}$) free to vary.\footnote{The temperature measured in the neutron star frame was converted to that of an observer in infinity via $kT^{\infty}_{\mathrm{eff}}= kT_{\mathrm{eff}} (1+z)^{-1}$, where $(1+z) = 1/ \sqrt{1-2GM/(Rc^2)}$ is the gravitational redshift. Since we perform our spectral fits for $M=1.6~\Msun$ and $R=11$~km, we used $(1+z)=1.33$.} 
For the \textsc{pegpwrlw} model, the energy boundaries were set such that the model normalization represents the unabsorbed power-law flux in the 0.5--10 keV band. Furthermore, as the power-law index was poorly constrained in our data set, we fixed this parameter in all our spectral fits (see Section~\ref{subsec:qspec}). 

To obtain our final results, all quiescent \chan\ and \swift\ spectra were fitted simultaneously with the hydrogen column density ($N_{\mathrm{H}}$) tied between all epochs. The total and thermal 0.5--10 keV unabsorbed fluxes and errors were determined with the \textsc{cflux} model. All errors quoted in this work reflect 1$\sigma$ confidence levels.


\section{Analysis and results}

\subsection{Quiescent spectral evolution after the 2016 outburst}\label{subsec:qspec}
We first investigated the two individual \chan\ observations. For both observations one-component fits with an \textsc{nastmos} model leave strong residuals at energies above 3~keV, leading to unacceptable fits ($\chi_{\nu}^{2}$/dof $=2.55/46$ for observation 1 and $\chi_{\nu}^{2}$/dof $=1.72/41$ for observation 2). Adding a power-law component with $\Gamma = 1.7$ (see next paragraph), significantly improved the fits ($\chi_{\nu}^{2}$/dof $=1.07/45$ for observation 1 and $\chi_{\nu}^{2}$/dof $=1.00/40$ for observation 2), demonstrating that an additional spectral component is statistically required to describe the data (with f-test probabilities of $3\times10^{-10}$ and $2\times 10^{-6}$, respectively).  

Leaving the power-law index free in the \chan\ fits resulted in $\Gamma=1.0 \pm 0.6$ for the first spectrum, while this parameter was completely unconstrained for the second. Studying the decay of the bright 2010 outburst of \source\ with \chan\ and \xmm, \citet{campana2014} showed that the power-law index was consistent with being constant at a value of $\Gamma = 1.7\pm0.1$. It is not clear if the power-law index should be the same during the outburst decay and in quiescence, nor if the power-law index should be constant in quiescence. Various studies of \source\ in quiescence report a range of power-law indices \citep[$\Gamma$$\sim$0.8--2.7; e.g.][]{cackett2011_aqlx1,marino2018}. In the \swift\ quiescent study of \citet{waterhouse2016}, all spectra were fitted with $\Gamma = 1.7$ fixed, based on the results of \citet{campana2014}. To allow for a direct comparison with the outbursts studied in \citet{waterhouse2016}, we choose to fix the power-law index to the same value. To test if this choice affects our conclusions, we also carried out fits for $\Gamma = 1.0$ (i.e. the value obtained for our first \chan\ spectrum). 

\begin{figure}
 \begin{center}
\includegraphics[width=8.5cm]{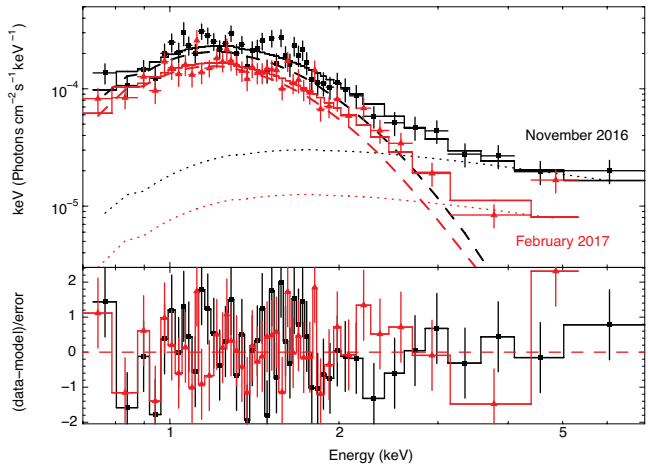}
    \end{center}
\caption[]{Unfolded \chan/ACIS-S spectra of \source\ obtained in 2016 November (black) and 2017 February (red). The spectral data are fitted to a two-component model (solid curves) comprised of a neutron star atmosphere (dashed curves) and a power law (dotted curves). The bottom panel shows the 1-$\sigma$ residuals of the fits. }
 \label{fig:chanspec}
\end{figure}

In the $\sim3$ months separating our two \chan\ observations, the temperature of the neutron star atmosphere decreased from $117.5 \pm 1.3$~eV to $110.7 \pm 1.1$~eV, and the corresponding thermal flux decreased by $\sim$30\% (see Table~\ref{tab:spec}). The flux in the power-law spectral component also decreased in strength during this time. As a result, the fractional contribution of this hard emission tail to the overall unabsorbed 0.5--10 keV flux is similar for the two observations; $\sim$20\% in the first and $\sim$14\% in the second. Figure~\ref{fig:chanspec} shows the two \chan\ spectra. 

\begin{figure}
 \begin{center}
\includegraphics[width=8.5cm]{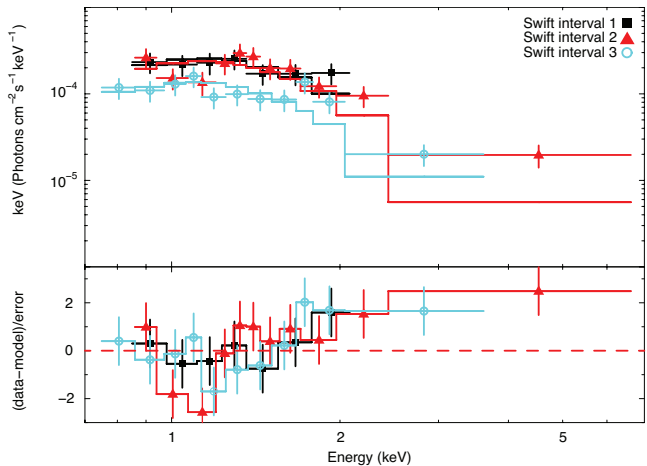}
    \end{center}
\caption[]{Unfolded \swift/XRT spectra of \source\ obtained in 2016 October (black squares), 2016 November (red triangles), and 2017 March--May (cyan circles). The spectral data are fitted to an absorbed neutron star atmosphere model (solid curves), to illustrate that this cannot describe the data at energies $>$2 keV (as shown by the 1-$\sigma$ fit residuals in the bottom panel). }
 \label{fig:swiftspec}
\end{figure}

When using an absorbed \textsc{nsatmos} model to describe the XRT spectra, we find excess emission above the model fit at energies $\gtrsim$2~keV, most notably for the second and third \swift\ epoch. This is shown in Figure~\ref{fig:swiftspec} and suggests that a hard emission tail is present in the XRT spectra as well. We therefore included a power-law component, with a fixed index of $\Gamma = 1.7$ and variable normalization, when fitting the XRT data.

Table~\ref{tab:spec} summarizes the results of fitting the three \swift/XRT data sets together with the two \chan\ spectra. The temperature of the neutron star atmosphere is observed to decrease from $118.9 \pm 4.5$~eV in 2016 October to $102.6 \pm 2.5$~eV in 2017 March--May. The 0.5--10 keV flux of the neutron star atmosphere is observed to decrease accordingly. All spectra are dominated by the neutron star atmosphere component, which has a fractional contribution to the total 0.5--10 keV flux of $\sim$70\% to 90\%. 

\begin{figure}
 \begin{center}
\includegraphics[width=8.5cm]{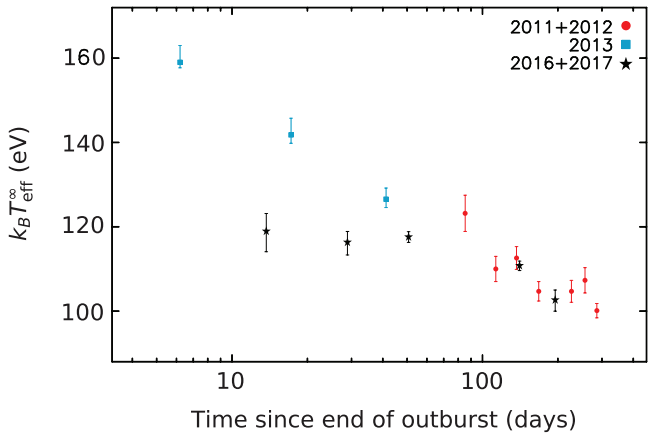}
    \end{center}
\caption[]{Evolution of the inferred neutron star temperature (for an observer at infinity) after the 2016 outburst (black stars) studied in this work, compared to that observed after the 2011 and 2013 outbursts reported by \citet{waterhouse2016}. The third and fourth data point obtained in 2016 are from \chan\ observations, whereas all other data points in this plot have been obtained from \swift/XRT observations.
}
 \label{fig:comparison}
\end{figure}

Looking at the temperature evolution in more detail, we see that during the first $\sim$50 days (covered by the first two \swift\ epochs and the first \chan\ observation) the neutron star temperature does not strongly change, while there is clear decrease in the $\sim$50 days separating the two \chan\ observations. As can be seen in Table~\ref{tab:spec} and Figure~\ref{fig:comparison}, the temperature had decreased further by the time of the last \swift\ epoch. 

In Figure~\ref{fig:comparison} we compare the temperature evolution inferred from our \chan\ and \swift\ analysis for the 2016 outburst with that observed by \swift/XRT after the 2011 and 2013 outbursts \citep[from][]{waterhouse2016}, which were analysed in the same way as we do here. The comparison with the 2011 (cyan squares) and 2013 (red filled circles) outbursts is of particular interest because these had such similar properties as the 2016 outburst that we study here (see Sections~\ref{subsec:swiftlc} and~\ref{subsec:ob}). It is immediately clear that at early times the absolute temperatures after the 2016 outburst are lower, and the temperature evolution is flatter, than seen after the 2013 outburst. There is thus clearly a difference in quiescent behavior after the two outbursts, as is also illustrated by the \swift/XRT quiescent light curves (Section~\ref{subsec:swiftlc}). It is worth emphasizing that the entire {\it shape} of the cooling curve is different, i.e. there is not simply a systematic shift in temperature between the different years.

Modeling all spectra with a power-law component of $\Gamma =1.0$ yields a systematic upward shift in temperature by $\sim$1--2~eV, which is well within the typical $1\sigma$ errors that we obtain (see Table~\ref{tab:spec}). Moreover, our main conclusions are based on the relative temperature evolution (i.e. the shape of the cooling curve), in particular in comparison with the 2013 data, which are unaffected by this systematic shift, hence our choice of $\Gamma$. 
Having measured the temperature evolution of \source\ following the end of its 2016 outburst, we proceed by modeling these data with a thermal evolution model to put physical constraints on the properties of the crust.

\begin{table*}
\caption{Spectral analysis results for 2016--2017 quiescence observations of \source.\label{tab:spec}}
\begin{threeparttable}
\begin{tabular*}{1.01\textwidth}{@{\extracolsep{\fill}}lccccccc}
\hline
Instrument & Epoch & mean MJD & ObsID(s) & $kT^{\infty}$  & $F_{\mathrm{th}}$ & $F_{\mathrm{X}}$  & $L_{\mathrm{X}}$   \\
& &  &  & (eV) & \multicolumn{2}{c}{($\times10^{-12}~\flux)$} & ($\times10^{33}~\lum)$ \\
\hline
\swift/XRT & 2016 epoch 1 & 57677.7 & 34719016--20& $ 118.9\pm4.5$ & $1.15 \pm 0.20$ & $ 1.46\pm0.30$ &$ 4.4\pm 0.9$    \\
\swift/XRT & 2016 epoch 2 & 57692.9 & 34719021--26& $ 116.2\pm2.8$ & $0.97 \pm 0.14$ & $ 1.17\pm 0.19$ &$ 3.5\pm 0.6$   \\
\chan/ACIS & 2016 Nov 22 & 57714.8 & 18984 & $117.5 \pm 1.3$ & $1.10 \pm 0.13$ & $ 1.50\pm 0.11$ &$ 4.5\pm 0.3$   \\
\chan/ACIS & 2017 Feb 20 & 57804.7 & 18985 & $110.7\pm1.1$ & $0.81 \pm 0.09$ & $ 0.98\pm 0.06$ &$ 2.9\pm 0.2$   \\
\swift/XRT & 2017 epoch 3 & 57859.1 & 33665090--99 & $102.6 \pm 2.5$ & $0.62 \pm 0.38$ & $ 0.82\pm 0.08$ &$ 2.5\pm 0.3$    \\
\hline
\end{tabular*}
\begin{tablenotes}
\item[]Note. -- 
$F_{\mathrm{th}}$ is the unabsorbed flux from the neutron star atmosphere, $F_{\mathrm{X}}$ is the total unabsorbed flux and $L_{\mathrm{X}}$ is the total luminosity assuming a distance of 5~kpc (all in the energy band of 0.5--10 keV). The joint spectral fit resulted in $N_{\mathrm{H}}$$=$$(6.6 \pm 0.3) \times 10^{21}~\nh$ and $\chi_{\nu}^2=0.98$ for 139 dof. The following parameters were kept fixed in the fits: $\Gamma$$=$$1.7$, $M$$=$$1.6~\Msun$, $R$$=$$11$~km, $D$$=$$5$~kpc, $N_{\mathrm{nsatmos}}=1$. Errors represent 1$\sigma$ confidence intervals.
\end{tablenotes}
\end{threeparttable}
\end{table*}

\subsection{Properties of the 2016 outburst}\label{subsec:ob}
Since the energy released in the crust of neutron stars due to nuclear reactions is proportional to the mass-accretion rate \citep[e.g.][]{haensel2008,steiner2012}, the outburst properties have to be taken into account in crust cooling simulations \citep[e.g.][]{brown08,ootes2016}. 
\citet{ootes2018} calculated the long-term (1996--2015) bolometric flux light curve of \source\ from \rxte, \maxi, and \swift\ data by using instrument-specific count-rate conversion factors for hard and soft spectral states (see their Table~1). Applying the same method to calculate the energetics of the 2016 outburst yields an average mass-accretion rate of $3.2 \times 10^{-9}~\mdot$. This is $\sim$20--30\% higher than the values obtained for the 2011 and 2013 outbursts \citep[$2.7\times10^{-9}$ and $2.3\times10^{-9}~\mdot$, respectively;][]{ootes2018}. 


Another ingredient in crust cooling simulations that can, in principle, be constrained from observations is the amount of light elements present in the accreted envelope. This determines how the observed effective temperature maps on to the actual interior temperature. As mass is accreted, the amount of light elements in the envelope increases but it suddenly drops when an X-ray burst occurs (since the light elements are then fused into heavier elements). We did not find any reports of X-ray bursts during the 2016 outburst in the literature, nor are there any X-ray bursts detected in the \swift/XRT observations. However, since \source\ is known to display X-ray bursts regularly \citep[e.g.][]{galloway06}, it is very likely that these were simply missed due to limited sampling and observing time. This implies that we cannot determine the last instance at which an X-ray burst occurred during the 2016 outburst, which would provide an upper limit on the He column depth accumulated before the start of quiescence (because it is plausible that a later X-ray burst was missed, which would lower the amount of light elements). We therefore left this parameter free in our crust cooling simulations (Section~\ref{subsec:coolmodel}).

\subsection{Crust cooling simulations}\label{subsec:coolmodel}
Based on the properties we infer for the 2016 outburst (Section~\ref{subsec:ob}), and taking into account the outburst history of \source, we modeled the temperatures obtained in the subsequent quiescent phase with the thermal evolution code \textsc{NSCool} \citep[][]{page2016_nscool,ootes2016,ootes2018}. To be consistent with our spectral fits, we assumed $M=1.6~\Msun$ and $R=11$~km in all thermal evolution simulations. The impurity parameter of the crust ($Q_{\mathrm{imp}}$) was assumed to be 1, since \citet{ootes2018} did not find any evidence of a higher impurity parameter when modeling the long-term outburst and quiescence data of Aql X-1. This is not surprising, since at the high crust temperatures found for \source, electron--impurity scatterings do not influence the thermal conductivity \citep[see for instance figure 9 of][]{page2012}.

Modeling the outburst (and cooling) history is particularly important for \source\ because the recurrence time of the outbursts is too short to allow the crust to fully cool \citep[][]{ootes2018}. In other words, the observed crust cooling behavior seen after an outburst depends on the crust temperature profile when the outburst commences, and in case of \source\ this depends on the accretion history. In first instance, we ran a fit keeping all parameters for the previous outbursts fixed at the values found by \citet{ootes2018}, including the core temperature in the neutron star frame ($T_{0}=8.9\times10^{7}$~K). In this simulation therefore only the envelope composition ($Y_{\mathrm{C}}$), shallow heating strength ($Q_{\mathrm{sh}}$), and shallow heating depth ($\rho_{\mathrm{sh,min}}$) were free to vary for the 2016 outburst. 

\begin{figure}
 \begin{center}
\includegraphics[width=8.5cm]{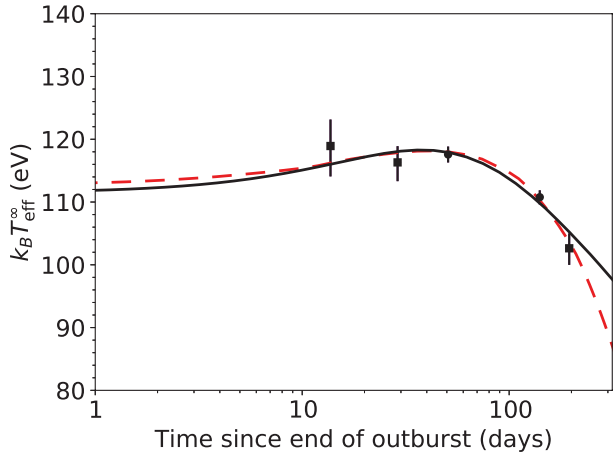}
    \end{center}
\caption[]{Neutron star temperatures inferred from the 2016--2017 spectral data of \source\ compared to thermal evolution simulations. The \swift\ data are shown as squares, the \chan\ data as circles. The black solid curve indicates our standard model, the red dashed curve a model with a colder core (see Section~\ref{subsec:coolmodel} for details).
}
 \label{fig:sim}
\end{figure}

The results of our simulations are listed in Table~\ref{tab:sim} and shown as the black solid curve in Figure~\ref{fig:sim}. We find that reproducing the data requires a large amount of shallow heat during the 2016 outburst, located relatively deep in the crust. The magnitude of shallow heating is a factor $\sim$2.5--4 higher, and the depth is a factor $\sim6$ larger, than obtained for the 2011 and 2013 outbursts of \source\ \citep[][]{ootes2018}, which are listed in Table~\ref{tab:sim} for comparison.\footnote{It is of note that due to the lack of data within the first $\sim 100$~days of quiescence (see Figure~\ref{fig:lcxrt}), the shallow heating parameters for the 2011 outburst are less well constrained than for the 2013 and 2016 outbursts.} The fitted light envelope composition is not strongly constrained, but the obtained values are similar for the three outbursts.

The main observational features that determine our derived shallow heating parameters are the flatness of the cooling curve in the first 50 days of quiescence, and the temperature drop that is observed after this. Indeed the reason that, in the model for the 2016 outburst, the shallow heating must be located very deep, results from the fact that the cooling starts relatively late.\footnote{For comparison, in \maxisource\ for instance, the strong temperature drop that constrains the depth of the shallow heating is observed at $\sim$10--20 days after the outburst \citep[][]{parikh2017_maxi}.} Moreover, the inferred strength of the shallow heating is driven by the relatively strong temperature drop observed at this time. We note that although the inferred strength of the shallow heating is higher for the 2016 outburst than for the 2013 one, the temperature observed shortly after the outburst is lower for 2016, because the shallow heating is located much deeper (see Table~\ref{tab:sim}). 

Seeking alternative ways to explain the sudden temperature drop at $\sim 100-150$~days into quiescence, we considered the possibility that \source\ might have a very cold core. As pointed out by \citet{ootes2018}, we likely have never observed \source\ at its true base level because the crust does not have time to fully cool in between outbursts. We therefore performed another run in which we fit for the core temperature in addition to the envelope composition and shallow heating parameters. The results are listed in Table~\ref{tab:sim} and shown as the red dashed curve in Figure~\ref{fig:sim}. Since a lower core temperature results in a stronger temperature gradient, hence stronger cooling, the amount of required shallow heating is reduced. However, we still need a significant amount of shallow heating ($\sim 5$~MeV~nucleon$^{-1}$) to explain the observed quiescent temperature evolution. The requirement for having this heat located relatively deep in the crust also remains, since it is determined by the late time of the temperature drop.

We emphasize that the need for strong shallow heating, relatively deep in the crust, is driven by the shape of the cooling curve and not the absolute temperatures. As stated in Section~\ref{subsec:qspec}, the temperatures inferred from spectral fitting shift $\sim1-2$~eV upward if a power law with $\Gamma=1.0$, instead of $\Gamma=1.7$, is used. This small systematic increase in temperature does not affect our conclusions about the depth and strength of shallow heating. 

\begin{table}
\caption{Results of the crust cooling simulations.\label{tab:sim}}
\begin{threeparttable}
\begin{tabular*}{0.49\textwidth}{@{\extracolsep{\fill}}lccc}
\hline
Outburst & log $Y_C$ & $Q_{sh}$ & $\rho_{\mathrm{sh,min}}$    \\
& & (MeV~nucleon$^{-1}$)  & ($\dens$)   \\
\hline
2016 & $6.6^{+0.9}_{*}$ & $9.2 \pm 1.6$  & $2.8^{+0.1}_{-0.2}\times 10^{10}$   \\
2016 - cold core & $10.5^{+0.3}_{-0.8}$ & $5.3^{+5.4}_{-0.7}$ & $3.4^{+1.2}_{-0.6}\times 10^{10}$  \\
2013 & $8.8^{+1.1}_{-1.5}$ &  $2.3^{+0.5}_{-0.3}$ & $0.4^{+0.7}_{*}\times 10^{9}$   \\
2011 & $8.3^{+0.7}_{-0.9}$ & $3.7^{+1.5}_{-0.9}$  & $0.4^{+7.9}_{*}\times 10^{9}$  \\
\hline
\end{tabular*}
\begin{tablenotes}
\item[]Note. -- 
The quiescent data obtained after the 2016 outburst are fitted in this work. The quoted values for the 2011 and 2013 outburst were obtained by \citet{ootes2018} and are listed here for comparison. Errors represent 1$\sigma$ confidence levels. An asterisk is used whenever the error range hit the hard boundary of that fit parameter in the model. 
In the standard fit for the 2016 data, the core temperature in the neutron star frame was fixed to the value obtained by \citet{ootes2018}, $T_{0}=8.9\times10^{7}$~K. In the alternative fit we allowed the core temperature to be lower, which yielded $T_{0}=2.9^{+1.0}_{-0.6} \times10^{7}$~K. 
\end{tablenotes}
\end{threeparttable}
\end{table}


\section{Discussion}\label{sec:discuss}
We report on \chan\ and \swift\ observations obtained after the bright 2016 outburst of \source\ ended. The aim was to i) further test the hypothesis that the bulk of the quiescent flux evolution of \source\ is driven by cooling of an accretion-heated crust, and ii) based on this assumption, better constrain the properties of the neutron star crust. In particular, we are interested in gaining more insight into the nature of the puzzling source of shallow heating that has been inferred for several neutron star LMXBs, including \source, in the crust-cooling hypothesis \citep[e.g.][]{page2013,degenaar2014_exo3,degenaar2015_ter5x3,parikh2017,parikh2018,ootes2016,ootes2018}. If crust cooling can be studied in \source, its frequent outbursts provide the opportunity to study the properties of this shallow heating after several different outbursts, thereby allowing to break degeneracies with neutron star specific parameters that may be involved in shallow heating and do not change between different outbursts \citep[e.g. mass, spin, magnetic field strength;][]{waterhouse2016}.

There are two other neutron stars for which crust cooling has been studied after different outbursts; MAXI J0556--332 \citep[][]{homan2014,parikh2017_maxi} and MXB 1659--298 \citep[e.g.][]{wijnands2004,cackett2013_1659,parikh2018}. For MAXI J0556--332, which exhibited 3 outbursts of different duration and peak intensity, there is clear evidence that the depth and magnitude of shallow heating varies between the different outbursts. This rules out that basic neutron star parameters play an important role in regulating shallow heating \citep[][]{parikh2017_maxi}. MXB 1659--298, on the other hand, showed remarkably consistent heating parameters for 2 outbursts that were of similar brightness but different duration \citep[2.5 and 1.5~yr;][]{parikh2018}. In the case of \source, it was previously noted by \citet{waterhouse2016} that the 2011 and 2013 outburst properties were very similar, and that the quiescent \swift/XRT count rate light curves and temperature evolution also gave consistent results between the two outbursts. Indeed, the detailed study of \citet{ootes2018} yielded similar shallow heating parameters for the 2011 and 2013 outbursts (see also Table~\ref{tab:sim}).\footnote{A note of caution is that there is only short time overlap in the data; for the 2013 outburst there is only quiescent coverage up to $\sim 100$~days into quiescence, whereas monitoring after the 2011 outburst did not start until this point (see Figure~\ref{fig:lcxrt} and~\ref{fig:comparison}).} 

Based on X-ray burst studies of the persistently accreting neutron star LMXB 4U 1820--30, it was previously suggested that differences in the accretion geometry could perhaps lead to different levels of shallow heating \citep[][]{zand2012}. \source\ is a promising target to test this idea, because it exhibits different classes of outbursts that have different spectral state evolution and hence likely different accretion geometries \citep[e.g.][]{maitra2008,asai2012}. In modeling multiple outbursts of \source, \citet{ootes2018} tested if shallow heating could be operational only in a particular spectral state (hard or soft), but ruled out such a simple connection. The data that we present here provides another opportunity to test any possible link between shallow heating and spectral states. 
The \maxi\ and \swift\ (BAT and XRT) light curves of the 2016 outburst of \source\ are remarkably similar to that of the bright 2011 and 2013 outbursts (see Figures~\ref{fig:lcxrt} and~\ref{fig:lcmonit}). This makes it likely that the spectral state behavior (i.e. the accretion geometry) was very similar during the three outbursts \citep[which was also noted by][]{diaztrigo2018}. If the global outburst properties are the main driver in determining the shallow heating, we would therefore expect to see similar crust cooling behavior after the three outbursts. 

The observed early thermal evolution after the 2016 outburst was, however, markedly different than seen after the 2013 one. The neutron star temperature was much lower right after the outburst ended, and it did not strongly evolve until $\sim$100 days into quiescence, when a decrease in temperature is observed. By using thermal evolution simulations, we determined that, within our current understanding of heating and cooling models, the late temperature drop seen after the 2016 outburst can only be achieved if the shallow heating was much stronger, and located much deeper in the crust, than for the 2013 outburst. Our results imply that the spectral state behavior during an outburst (i.e. the accretion geometry) cannot play a strong role in setting the properties of shallow heating.

\subsection{Implications for our understanding of shallow heating}\label{subsec:implications}
Whereas the concept of shallow heating has been known for over a decade now \citep[e.g.][]{brown08}, its physical origin is still not understood. Nevertheless, ongoing crust cooling studies are providing us with pieces of information regarding the mechanism of shallow heating. In particular, the studies of crust cooling of 3 sources that exhibited multiple outbursts (MAXI J0556--332, MXB 1659--28 and \source) has established that shallow heating does not strongly depend on i) neutron star parameters that do not change between outbursts \citep[this work and][]{parikh2017_maxi}, and ii) the accretion geometry \citep[this work and][]{ootes2018}. Therefore, there must be other factors that drive the shallow heating. 

One framework in which different shallow heating for similar types of outbursts can be accommodated, is that of chemical convection. In this model, the mixture of elements left in the neutron star envelope after thermonuclear burning organizes itself into layers of light and heavy elements \citep[][]{horowitz2007}. This chemical separation may drive a convective heat flux that can potentially heat the outer layers of the crust \citep[][]{medin2011,medin2014,medin2015}. Depending on the last instance at which an X-ray burst occurred before accretion switched off, the envelope composition can differ, even if the outbursts are very similar \citep[][]{brown2002}. Chemical convection will act differently depending on the envelope composition and could therefore lead to different shallow heating. However, it can likely generate only a few tenths of an MeV of energy per accreted nucleon \citep[][]{medin2015}, and therefore cannot account for the strong shallow heating ($\sim 5-10$~MeV~nucleon$^{-1}$, depending on the core temperature) that we infer for the 2016 outburst of Aql X-1. Similar conclusions were drawn for the strong shallow heating inferred for the main outburst of MAXI J0556--332 \citep[][]{deibel2015,parikh2017_maxi}. 

One source of energy that can potentially provide the strong shallow heating inferred for \maxisource\ and \source, is the orbital energy of the accreted material \citep[][]{inogamov2010}. It was previously suggested, for \maxisource, that dissipation of accretion-generated oscillations (``g modes'') in the liquid part of the neutron star crust (the ``ocean'') could inject heat in the crust \citep[][]{deibel2015,deibel2016}. However, since the observed properties of 2016 outburst of \source\ were so similar to that of its 2011 and 2013 outbursts, it seems difficult to understand why the heat injected in the crust would not be similar. This is, at least, under the assumption that the X-ray properties are a reasonable proxy for the mass-accretion rate onto the neutron star, which may not necessarily be true \citep[e.g.][]{vanderklis2001}.

Our results for the 2016 outburst of \source\ do not only stand out from its 2013 (and 2011) outburst, but also from the other crust cooling sources. Indeed, the magnitude of the shallow heating that we require is much higher ($\sim5-10$~MeV~nucleon$^{-1}$) than inferred for most other crust cooling sources, which typically require 1--2~MeV~nucleon$^{-1}$ \citep[e.g.][]{page2013,degenaar2014_exo3,degenaar2015_ter5x3,parikh2017,parikh2018,ootes2016,ootes2018}, except for MAXI J0556--332 during its main outburst, for which $\sim15-17$~MeV~nucleon$^{-1}$ was inferred \citep[][]{deibel2015,parikh2017_maxi}. The depth of shallow heating that we infer for \source, $\sim10^{10}$~g~cm$^{-3}$, is also higher than the typical values inferred for other crust cooling sources (using \textsc{NSCool}); these are often on the order of $10^8-10^9$~g~cm$^{-3}$ \citep[e.g. for \ks\ and \rxs;][]{ootes2016,parikh2017}, although sometimes the constraints are poor and much higher densities of $\sim10^{10}$~g~cm$^{-3}$ are allowed \citep[e.g. for \mxb\ and the second outburst of \maxisource;][]{parikh2017_maxi,parikh2018}. 

Although the shallow heating depth for the 2016 outburst is high, it is not excessive. For example, for the main outburst of \maxisource, which also required the very strong shallow heating, the inferred depth was $\sim 5\times 10^9$~g~cm$^{-3}$ \cite[][]{parikh2017_maxi}. We also note that this depth is still well below that of deep crustal heating, which occurs at densities of $10^{12}-10^{13}$~g~cm$^{-3}$ \citep[e.g.][]{haensel2008}. \citet{steiner2012} proposed that uncertainties in the nucleon symmetry energy could allow for much stronger deep crustal heating ($\sim5$~MeV~nucleon$^{-1}$) than previously assumed ($\sim2$~MeV~nucleon$^{-1}$), although more detailed calculations are needed to confirm this \citep[][]{fantina2018}. Moreover, it is unlikely that a different amount of deep shallow heating can account for our results of \source, as it acts at so much higher densities and would thus manifest itself at much later time in the cooling curve \citep[not probed by our observations;][]{brown08}.

We can speculate that how, or when, shallow heating is operating, somehow depends on the initial temperature of the crust when an outburst commences. The outburst that preceded the 2016 activity of \source\ was shorter and fainter than the outbursts that occurred before the 2011 and 2013 ones \citep[see][]{ootes2018}. Since the crust cooling time of this neutron star is shorter than the quiescent time between two outbursts, the initial temperature of the crust can significantly differ between outbursts, and perhaps this can explain the different levels of shallow heating inferred. 
A similar argument might apply for \maxisource; when its second and third outburst occurred, the crust had not yet relaxed from the strong heating of its first (main) outburst \citep[][]{parikh2017_maxi}. Perhaps the much higher crust temperature at the start of the second and third outburst is related to the fact that the shallow heating inferred for these outbursts was very different from that obtained for the first outburst \citep[$\lesssim2$ and $\sim17$~MeV~nucleon$^{-1}$, respectively;][]{parikh2017_maxi}. For MXB 1659--29, on the other hand, the crust temperature at the start of its last two outbursts was likely similar \citep[]{parikh2017}, and perhaps that can explain why there was no apparent difference in shallow heating for this source.\footnote{Although the quiescent phases preceding the 1999--2001 and 2015--2017 outbursts differed by a few years, the quiescent temperature evolves only slowly after a decade. It is therefore likely that the crust temperature was similar before both outbursts.}

\subsection{Other explanations for the quiescent evolution of Aql X-1}\label{subsec:accretion}
Other than crust cooling, the quiescent flux evolution of \source\ could possibly be powered by ongoing low-level accretion \citep[e.g.][]{kuulkers2008,cotizelati2014}. Nevertheless, in this interpretation it may not obvious either why the quiescent evolution of \source\ after its 2016 outburst should be significantly different from that observed after the 2013 outbursts. The similarity in outburst properties suggests that a similar part of the disk was involved in the outburst, and that a comparable amount of mass drained from the disk on to the neutron star. Therefore, one might expect a similar quiescent evolution, opposed to what is seen. 

In assessing the low-level accretion scenario for \source, it is worth noting that the fractional contribution of the hard emission component to the 0.5--10 keV unabsorbed flux is generally low; $\lesssim$20\% in that data that we presented here and in \citet{waterhouse2016}. \citet{wijnands2015} proposed that if the power-law and thermal component both contribute $\sim$50\% to the total quiescent flux, the emission is likely powered by accretion, while a lower power-law contribution may point to a different origin. Indeed, the most proximate neutron star LMXB, Cen X-4, exhibits equal flux contributions of its two quiescent emission components \citep[e.g.][]{cackett2010_cenx4}, and there is strong evidence that both are powered by low-level accretion onto the neutron star surface \citep[][]{bernardini2013,cackett2013_cenx4,chakrabarty2014,dangelo2014}. For \source, the power-law contribution is generally lower than $\lesssim$50\% \citep[e.g.][]{marino2018}. Nevertheless, the fractional contribution of the power-law remained approximately constant between our two \chan\ observations, i.e. the neutron star atmosphere and the power-law component decreased in tandem. This suggests that the two components are connected, possibly arising from the same emission process, which could be more naturally explained by continued accretion \citep[e.g.][]{cackett2010_cenx4}.

Other than turning to a different explanation for the quiescent emission altogether, it is also possible that our results on \source\ are exposing that we are still missing something in our understanding of heating and cooling of neutron star crusts. For instance, as noted earlier, a different envelope composition would in principle produce a systematic temperature shift, because it yields a different mapping between the surface and interior temperature of the neutron star, and hence not explain the different shape of the 2016 cooling curve of \source. However, it was very recently shown that the envelope composition can potentially significantly change during quiescence as a result of diffuse nuclear burning, a process in which elements diffuse to such depths where the density and temperature are sufficiently high to ignite nuclear burning \citep[][]{wijngaarden2019}. The impact of this process on crust cooling studies, and if it can potentially explain the observed behavior of \source, needs to be explored.

\section*{Acknowledgements}
ND is supported by a Vidi grant awarded by the Netherlands organization for scientific research (NWO). LO, RW and AP are supported by an NWO top grant, module~1, awarded to RW. DP is partially supported by the Consejo Nacional de Ciencia
y Tecnolog{\'\i}aa with a CB-2014-1 grant \#240512. DA acknowledges support from the Royal Society. JMM and JH acknowledge support from \chan\ grant GO7-18031B. This work made use of data supplied by the UK \swift\ Science Data Centre at the University of Leicester. We also made use of MAXI data provided by RIKEN, JAXA and the MAXI team, and public light curves from the \swift/BAT transient project.

\footnotesize{
\bibliographystyle{mn2e}

}

\end{document}